\documentclass[prl,twocolumn,english,showpacs,superscriptaddress]{revtex4}
\usepackage[latin9]{inputenc}
\usepackage{amsmath,amssymb,amsfonts}
\usepackage{CJK}
\usepackage{bm}
\usepackage{tikz}
\usetikzlibrary{matrix,fit}

\usepackage{slashed}

\def\photonvex{\tikz{\fill[gray] (0,0) circle (3pt);} }
\def\chiralvex{\tikz{\fill[gray] (0,0) rectangle (6pt, 6pt);}}

\begin{document}

\title{  
Disordered Weyl semimetals and their topological family}

\author{Y. X. Zhao}
\email[]{yuxinphy@hku.hk}

\author{Z. D. Wang}
\email[]{zwang@hku.hk}

\affiliation{Department of Physics and Center of Theoretical and Computational Physics, The University of Hong Kong, Pokfulam Road, Hong Kong, China}

\date{\today}

\begin{abstract}
 We develop a topological theory for disordered Weyl semimetals in the framework of gauge invariance of replica formalism and boundary-bulk correspondence of Chern insulators.  An anisotropic topological $\theta$-term is analytically derived for the effective non-linear sigma model. It is this nontrivial topological term that ensures the bulk transverse transport 
 of Weyl semimetals to be robust against disorders. 
Moreover,  we establish a general diagram that reveals the intrinsic relations among topological terms in the non-linear sigma models and gauge response theories respectively for $(2n+2)$-dimensional topological insulators, $(2n+1)$-dimensional chiral fermions, $(2n+1)$-dimensional chiral semimetals, and $(2n)$-dimensional topological insulators with $n$ being a positive integer.
\end{abstract}

\maketitle

\textit{Introduction}
Recently, Weyl semimetal (WSM) has been attracting more and more attention due to its interdisciplinary interest from anomalous transport in condensed matter physics~\cite{WSM1,WSM2,WSM3,WSM4,WSM5,WSM6,WSM7,Fermi-arc,Experiment1,Experiment2} and quantum field theory anomalies as well as a  topological 
character of its gapless modes~\cite{Anomaly1,Anomaly2,Volovik-Book,Response0,Response1,Response2,Response3,Response4,Response5,Response6}. Weyl fermion as the quasi-particle of WSM has a definite chirality, left-handed or right handed, depending on its sign of spin polarization along the momentum direction\cite{Peskin-Book}. According to the no-go theorem\cite{No-go}, gapless Weyl points in a WSM appear as left and right-handed pairs in the momentum space. Since all the energy bands of a system with both time-reversal symmetry(TRS) and inversion symmetry(IS)  are doubly degenerate for both chiral modes, a WSM can only be realized in a system breaking TRS and/or IS symmetry. Here we focus essentially on the simplest model of WSM with two Weyl points, whose low-energy effective Hamiltonian reads
\begin{equation}
\mathcal{H}_{WSM}(\mathbf{k},\mathbf{b})=\begin{pmatrix}-\mathbf{\sigma}\cdot(\mathbf{k}-\mathbf{b}(\mathbf{x}))\\
 & \sigma\cdot(\mathbf{k}+\mathbf{b}(\mathbf{x}))
\end{pmatrix},\label{Model}
\end{equation}
where $\sigma$'s are Pauli matrices and $\mathbf{b}$ is the displacement of Weyl points from the origin of the momentum space. Under semi-classical approximation,  the spatial dependence of $\mathbf{b}(\mathbf{x})$ is assumed to be adiabatic. The model does not have TRS~\cite{note0}.  Each Weyl point has a nontrivial topological charge~\cite{Volovik-Book,FS-classification}, namely the Chern number on the gapped two-dimensional sphere enclosing a single Weyl point in $\mathbf{k}$ space is $\pm1$, which leads to topological terms in the $U(1)$ response of this model~\cite{Response0,Response1,Response2,Response3,Response4,Response5,Response6}. For real materials, disorders are normally unavoidable, and can be studied by the replica method, where a non-linear sigma model (NL$\sigma$M) encodes fluctuations of Nambu-Goldstone modes that are on a curved manifold\cite{Replica-I,Replica-II,Replica-III}.  Particularly for topological materials, the global topology of the target manifold may have essential effects, such that topological terms in the NL$\sigma$M are significantly important for describing physics of the system under disorders~\cite{Pruisken,Ryu-Z2,Supersymmetry-Z2,TI-Classification}, which highly motivates us to develop a general theory for disordered WSMs.

\begin{figure}
\centering
\includegraphics[scale=0.25]{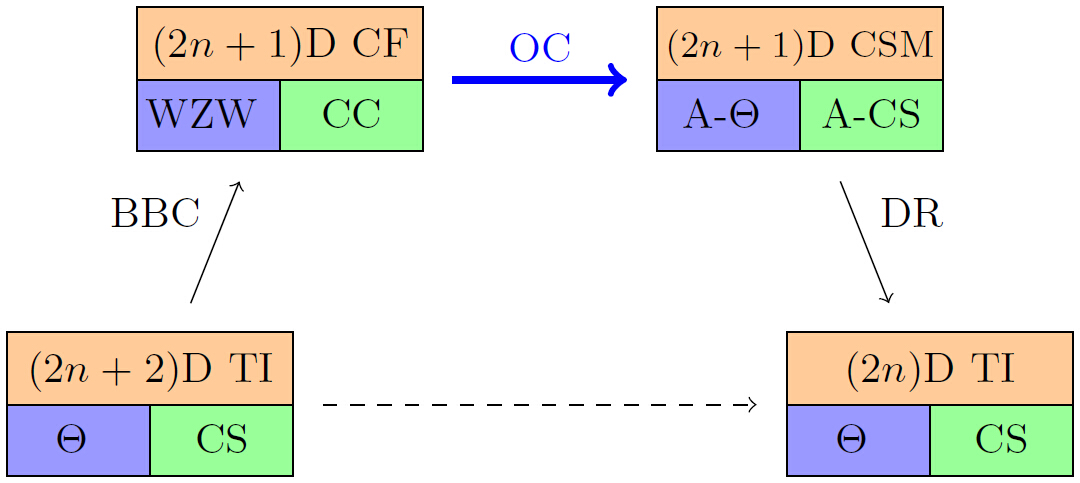}
\caption{ Relationship diagram of topological terms and materials for the $n$-th unit of a whole topological family of class A~\cite{Volovik-Book,FS-classification}. For each node, the upper, below-left, and below-right boxes denote the model, the topological term in NL$\sigma$M, and  gauge theory,  respectively.  The dashed arrow is effectively composed by the three successive solid arrows. Nodes: TI(topological insulator), $\Theta$($\theta$-term), CS(Chern-Simons term), CF(chiral fermion), WZW(WZW-term), CC(Chern character),   CSM(chiral semimetal), A-$\Theta$(anisotropic $\Theta$) and A-CS(anisotropic CS). Arrows: BBC(boundary-bulk correspondence), OC(opposite coupling) and DR(dimension reduction). Notably, the thick blue-arrow represents a key step that deduces the central result of this work. 
\label{diagram}}
\end{figure} 

In this Letter, for the model of Eq. (\ref{Model}) under disorders and preserving no discrete symmetry, we derive for the first time an anisotropic topological term in the action of the NL$\sigma$M  as
\begin{equation}
S_{A\Theta}=-\frac{1}{8\pi} \int d^3 x\,\, \epsilon^{ijk}b_i \mathrm{tr}(Q\partial_j Q\partial_k Q), \label{ATheta}
\end{equation}
where $\epsilon^{ijk}$ is the completely anti-symmetric tensor, $b_i$ denotes the $i$-th component of $\mathbf{b}$, and $Q$ stands for a sigma field to be specified later.  We refer to this term as an anisotropic $\theta$-term, as  it is just a usual $\theta$-term on the plane perpendicular to $\mathbf{b}$ for a constant $\mathbf{b}$~\cite{Pruisken}. Actually this term can be generalized to any chiral semimetal (CSM) with two chiral points separated in the $\mathbf{k}$ space of odd dimensions~\cite{note1}.  We also reveal that this new anisotropic $\theta$-term is associated with the so-called opposite coupling in CSM, which plays a key role in a relationship diagram of topological terms for a family of topological matter, as illustrated in Fig.(\ref{diagram}),
where each topological term in the NL$\sigma$M has a counterpart in the $U(1)$ gauge response theory.

\textit{Disorder model} We now consider that the WSM, Eq.(\ref{Model}), is subjected to a white-noise and random scalar potential $V(\mathbf{r})$, namely
 \begin{equation}
 \mathcal{H}=\mathcal{H}_{WSM}(\mathbf{b})+V(\mathbf{r}).
 \end{equation}
 Assuming that the probability density of $V$ is Gaussian, we have  $\overline{V(\mathbf{r})}=0$ and $\overline{V(\mathbf{r})V(\mathbf{r}')}=g^2\delta^{3}(\mathbf{r}-\mathbf{r}')$ with $g^2$ indicating the strength of the random potential.  After applying the replica scenario~\cite{Replica-I,Replica-II,Replica-III} and averaging over the random potential, the disordered system may be described by a Lagrangian
 \[
 \mathcal{L}=\psi_{a}^{\dagger}G^{-1}\psi_{a}-\frac{g^2}{2}\psi_{a}^{\dagger}\psi_{a}\psi_{b}^{\dagger}\psi_{b},
 \]
  where the spinor is decomposed into retarded and advanced space, namely $\psi=(\psi^r,\psi^a)^T$, and $G^{-1}=\omega-\mathcal{H}_{WSM}+i\eta\tau_{3}^{ra}$. The Pauli matrix $\tau^{ra}_3$ acts in this space, and $\eta$ is an infinitesimal positive number. The subscript as the replica index ranges from $1$ to $N$ with repeated one being summed over.  The four-operator term comes from the averaging over the random potential. Since our system is in the topological class A, we introduce the sigma field
\begin{equation}
Q\in\frac{U(2N)}{U(N)\times U(N)},
\end{equation}
to decouple the four-operator term~\cite{Pruisken,RMT-1,RMT-2}, 
\[
e^{-\frac{g^2}{2}\int \psi^\dagger_{\alpha}\psi^\dagger_{\beta}\psi_{\alpha}\psi_{\beta}}\sim \int \mathcal{D}Q\,\,e^{-\frac{\Delta^2}{4g^2}\int\mathrm{tr}Q^2+\Delta \int \psi^\dagger_\alpha Q_{\alpha\beta}\psi_\beta},
\]
where the Greek subscripts indicate both replica and retarded-advanced indexes. Accordingly the partition
function can be transformed to be in
the form,
\[
Z=\int\mathcal{D}Q\int\mathcal{D}\psi^{\dagger}\mathcal{D}\psi\:\exp(-\int d^{3}x\:\mathcal{L}_{f}+\frac{\Delta^{2}}{4g^2}\mathrm{tr}QQ),
\]
where
\[
\mathcal{L}_{f}=\psi_{a}^{\dagger}(G^{-1}-\Delta Q_{ab})\psi_{b}.
\]
$\Delta$ is determined by the consistence
equation $\pi/g^2=\mathrm{ln}[1+(a_{0}\Delta)^{-2}]$ with $a_{0}$
being the short-distance cutoff. Assuming that $\omega$ is small and $\Delta>0$, the dynamic terms of
the effective NL$\sigma$M are given by
\begin{eqnarray*}
S_{eff} & = & \mathrm{ln}\:\mathrm{Det}(\mathcal{H}_{WSM}+\Delta Q).
\end{eqnarray*}
It is observed that the operator in the determinant is diagonal in
the chirality space, recalling that $\psi_{\alpha}=(\xi_{\alpha},\chi_{\alpha})^{T}$, where $\xi$ and $\chi$ are Weyl spinors with opposite chirality. This means the effective NL$\sigma$M $S_{eff}$ for $\mathcal{H}_{WSM}$ is a summation of the NL$\sigma$Ms for the left and right-handed chiral components, namely
\begin{equation}
S_{eff}[Q]=S_{-,eff}[Q,-\mathbf{b}]+S_{+,eff}[Q,\mathbf{b}], \label{Sumrule}
\end{equation}
where the signs in front of $\mathbf{b}$ indicate that it couples oppositely to the left and right-handed ones. We below derive $S_{\pm,eff}$ separately.

\textit{Boundary-bulk correspondence}
Before deriving $S_{\pm,eff}[Q,\pm\mathbf{b}]$, let us introduce the Wess-Zumino-Witten term (WZW-term) for a single Weyl point at the center of $\mathbf{k}$ space with disorders but without any anti-unitary symmetry, from a viewpoint of boundary-bulk correspondence(BBC)~\cite{WZW-term}. For a $(2n+2)$-dimensional (D)  TI in the 
class A with a nontrivial Chern number $C$ in its bulk \cite{RMT-1,RMT-2}, there are $C$ flavors of chiral fermions with the same chirality on each of its $(2n+1)$D boundaries~\cite{TI-1,TI-2}.  The NL$\sigma$M of the $(2n+2)$D TI under disorders has a $\theta$-term~\cite{Pruisken}. In the infrared limit the behavior of the TI is entirely determined by its gapless boundary, and the coupling constant of the $\theta$-term becomes the Chern number $C$ under the renormalization flow~\cite{Pruisken}. To be concrete, for a 4D TI, the $\theta$-term of the NL$\sigma$M is given by
 \begin{equation}
 S_{\Theta}^{4D}=\frac{iC}{128\pi}\int d^{4}x\;\epsilon^{\mu\nu\rho\lambda}\mathrm{tr}Q\partial_{\mu}Q\partial_{\nu}Q\partial_{\rho}Q\partial_{\lambda}Q,\label{4DTheta}
 \end{equation}
  Viewing from the boundary of the TI, this implies that there exists a WZW-term at level $C$ in the NL$\sigma$M of the chiral fermions on the boundary~\cite{Pruisken,TI-Classification}. For the case of Eq.(\ref{4DTheta}), the corresponding WZW-term is found to be
  \begin{equation}
  S_{WZW}=\frac{i\nu_b}{128\pi}\int d\tau d^{3}x\;\epsilon^{\mu\nu\rho\lambda}\mathrm{tr}\tilde{Q}\partial_{\mu}\tilde{Q}\partial_{\nu}\tilde{Q}\partial_{\rho}\tilde{Q}\partial_{\lambda}\tilde{Q}\label{WZW},
  \end{equation}
where $\tau\in[0,1]$ is the extending parameter, and $Q(x)$ on the $S^3$ is extended continuously to $\tilde{Q}(x,\tau)$ with $\tilde{Q}(x,1)=Q(x)$ and $\tilde{Q}(x,0)$ being constant\cite{Note2}. In this case, $\tau$ can be regarded as the parameter of the radial direction if the geometry of the 4D TI is a disc $D^4$.  Here $\nu_b$ is the total $\mathbb{Z}$-type topological charge of the boundary gapless modes, which is defined as the Chern number on the gapped sphere enclosing these gapless points in momentum space,and it is equal to the bulk Chern number $C$ according to an established index theorem~\cite{Volovik-Book, FS-classification, FS-TI}. Inherently, the $\mathbb{Z}$-type topological charge  $\nu_b=C$ also implies the WZW-term to be at level $\nu_b$ in the NL$\sigma$M. Since this WZW-term has a topological character with a non-perturbative discrete coupling constant, gapless chiral modes have a topologically protected finite conductance, free from the Anderson localization~\cite{TI-Classification,Bosonization}. As the counterpart of the above results for NL$\sigma$M, the BBC can also be used to deduce the Chern character (CC) term in the gauge response of $(2n+1)$D chiral fermions from the Chern-Simons(CS) term of the gauge response of $(2n+2)$D TIs, since the CS term is not gauge invariant on a manifold with boundary, leading to the boundary CC terms.


\textit{The coupling of disordered Weyl fermions with a gauge field}
If $\mathbf{b}=0$, $S_{\pm,eff}(Q)$ just corresponds to chiral fermions under disorders.  According to the boundary-bulk correspondence, the NL$\sigma$M for a model $\mathcal{H}$ with only one gapless point contains a WZW-term,
\[
\Gamma_{m}[Q]=\frac{im}{128\pi}\int d\tau d^{3}x\;\epsilon^{\mu\nu\rho\lambda}\mathrm{tr}\tilde{Q}\partial_{\mu}\tilde{Q}\partial_{\nu}\tilde{Q}\partial_{\rho}\tilde{Q}\partial_{\lambda}\tilde{Q},
\]
where $m$ is the topological charge of the gapless modes in the model, raising from
 \[
 S[\chi,\chi^{\dagger}]=\int d^{3}x\;\chi^{\dagger}\left(\omega-\mathcal{H}+i\eta\tau_{3}^{ra}-V(\mathbf{r})\right)\chi .
 \]
 In the present case, $m=\pm$ correspond to the left and right-handed fermions, respectively. So it is clear that  the WZW-terms are cancelled if ever $\mathbf{b}=0$, leading to vanishing topological terms and  corresponding to a Dirac Hamiltonian under disorders.

 When $\mathbf{b}$ is nonzero, we can regard it as a chiral gauge field coupling oppositely to the left- and right-handed fermions. Let us consider $\mathcal{H}_{W,+}=\mathbf{\sigma}\cdot \mathbf{k}$ coupled with a $U(1)$ gauge field $\mathbf{A}$ under disorders,  noting that $\mathcal{H}_{W,-}=-\mathbf{\sigma}\cdot \mathbf{k}$ can be treated similarly. After application of the replica method, the Lagrangian  reads
  \[
 \mathcal{L}_{+}=\chi_{a}^{\dagger}(\omega\delta_{ab}+i\mathbf{\sigma}\cdot(\nabla+i\mathbf{A})\delta_{ab}+i\eta\tau_{3}^{ra})\chi_{b}-\frac{g^2}{2}\chi_{a}^{\dagger}\chi_{a}\chi_{b}^{\dagger}\chi_{b},
 \]
It is found that the $U(1)$ gauge
transformation can be made for each component of $\chi$ independently,
namely the action is invariant under the transformation,
\[
\chi^{a,s}\longrightarrow\chi^{a,s}e^{-i\alpha_{a,s}(\mathbf{x})}\]
and
\[\mathbf{A}(\mathbf{x})\longrightarrow\mathbf{A}(\mathbf{x})-\nabla\alpha_{a,s},
\]
with the superscript `$s$' being the index of the retarded-advanced space. Accordingly in the non-linear sigma version, $Q^{(a,s)(a's')}\sim\psi^{\dagger a,s}\psi^{a',s'}$
transforms under the gauge transformation as
\[
Q\longrightarrow e^{i\alpha_{a,s}}Q^{(a,s)(a's')}e^{-i\alpha_{a',s'}}.
\]
In the absence of  gauge field, the NL$\sigma$M is
\[
S_{+,eff}[Q]=\frac{1}{\lambda_+}\int d^{3}x\:\mathrm{tr}(\partial_{j}Q\partial_{j}Q)+\Gamma_{+}[Q],
\]
where $1/\lambda_+$ determined by the microscopic details of the Weyl point is proportional to the longitudinal conductivity. Our strategy is to find the minimal coupling of the NL$\sigma$M with the $U(1)$ gauge field, which is invariant under the above gauge transformation. To simplify our calculation but without loss of generality, we consider
a specific gauge transformation as
\[
\mathbf{A}(\mathbf{x})\longrightarrow\mathbf{A}(\mathbf{x})-\nabla\alpha\tau_{3}^{ra}
\]
 and
 \[
 Q\longrightarrow e^{i\tau_{3}^{ra}\alpha}Qe^{-i\tau_{3}^{ra}\alpha},
\]
which are the opposite constants in advanced and retarded spaces, respectively. Following a seminar work on current algebra of Witten~\cite{WZW-term}, we highlight our derivations below. 
The infinitesimal variation of $Q$ is
\[
Q\longrightarrow Q+i\alpha[\tau_{3},Q].
\]
$\tau_{3}$ operates on retarded and
advanced spaces, and hereafter we drop the superscript for simplicity. Since the minimal coupling for the ordinary term is readily obtained by the substitution, $\partial_j\longrightarrow\hat{\partial}_j=\partial_j+[iA_j,\,\,]$, we thus focus only on the WZW-term.
The variation of $\Gamma_{+}$ under the infinitesimal local
gauge transformation is
\[
\delta \Gamma_{+} = -\frac{1}{16\pi}\int_{S^{3}}\;\mathrm{tr}(\mathbf{d}\alpha\tau_{3\wedge}Q_{\wedge}\mathbf{d}Q_{\wedge}\mathbf{d}Q),
\]
where the exterior derivative and  wedge product have been used for brevity. Thus we have
\[
J_{WZW}=-\frac{1}{16\pi}Q\mathbf{d}Q_{\wedge}\mathbf{d}Q.
\]
We may expect that the coupling takes the form $\int\mathrm{tr}(\mathbf{A}_{\wedge}J)$. However, since the current $J_{WZW}$ is not gauge invariant, additional terms are needed to cancel its variation under gauge transformation, which turns out to be $\frac{1}{4\pi}\int\mathrm{tr}(\mathbf{A}_{\wedge}\mathbf{d}\mathbf{A}Q)$. As a result, the total gauge invariant action for the $\pm$ case is given by
\begin{eqnarray}
S_{\pm,eff}&=&\frac{1}{\lambda_{\pm}}\int d^{3}x\:\mathrm{tr}(\hat{\partial}_{j}Q\hat{\partial}_{j}Q)\pm\frac{1}{4\pi}\int\mathrm{tr}(\mathbf{A}_{\wedge}\mathbf{d}\mathbf{A}Q)\nonumber\\
&&\mp\frac{1}{16\pi}\int\mathrm{tr}(\mathbf{A}_{\wedge}Q\mathbf{d}Q_{\wedge}\mathbf{d}Q)+\Gamma_{\pm}. \label{EM}
\end{eqnarray}

\textit{Opposite coupling}
We are now ready to obtain the NL$\sigma$M of the WSM. In the WSM $\mathcal{H}_{+}$($\mathcal{H}_{-}$) is coupled with $+\mathbf{b}$($-\mathbf{b}$). Thus from Eqs.(\ref{Sumrule}) and (\ref{EM}), it is found  that
\begin{equation}
S_{eff}[Q]=\frac{1}{\lambda}\int d^{3}x\:\mathrm{tr}(\partial_{j}Q\partial_{j}Q)+S_{A\Theta}[Q], \label{Seff}
 \end{equation}
 where $1/\lambda=1/\lambda_++1/\lambda_-$, recalling that $S_{A\Theta}$, as our most main result,  is given by Eq.(\ref{ATheta}). Also note that the first term in Eq.(\ref{Seff}) is actually a normal action that accounts for usual non-topological properties.

A physical meaning of the anisotropic $\theta$-term becomes clear if $\mathbf{b}$ is constant. Intuitively, a 2D slice at any $\mathbf{k}\in(-\mathbf{b},\mathbf{b})$ along $\mathbf{b}$ direction may be viewed as a 2D TI with unit Chern number, which is accompanied with the bulk transverse conductivity  and the edge chiral gapless  modes (see the latter discussion around Eq.(\ref{2DTheta})).  
It is interesting  to note  that the transverse conductivity in the plane perpendicular to $\mathbf{b}$ is proportional to the magnitude of $\mathbf{b}$, being completely independent of the disorder strength $g$. The anisotropic form in the bulk is also consistent with the edge currents traveling perpendicular to $\mathbf{b}$, enabling them to have topological protection similar to that of the WZW-term. On the other hand, it is seen from the derivation that  the WZW-terms of two Weyl points are cancelled, which relies only on the fact that the two Weyl points have opposite topological charges, independent of their positions  in the $\mathbf{k}$-space. When the WZW-term is absent after the opposite coupling, its corresponding topological protection 
is lost, consistent with the fact that disorders may mix the two Weyl points leading to localization\cite{Inter-points}. However, 
remarkabally,
the new nontrivial anisotropic $\theta$-term emerges, such that it embodies the remaining anisotropic topological protection originated from topological charges after the global cancellation, analogous to the net electric field generated from an electric dipole.

\begin{figure}
\centering
\includegraphics[scale=0.3]{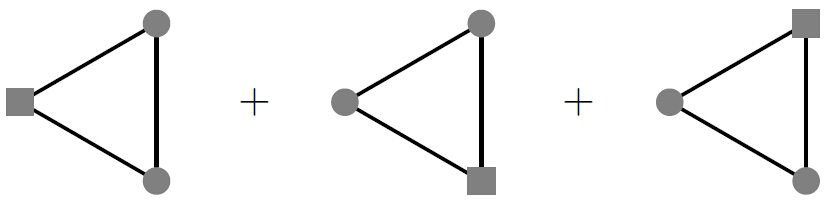}
\caption{Three-vertex diagrams\label{Feynman}}

\end{figure}

At this stage, we elaborate how to derive the A-CS term 
in Fig.(\ref{diagram}) 
(see Supplemental Material for details~\cite{supple}), 
from a trick of opposite coupling.
Treating $\mathbf{b}$ as a gauge field coupling oppositely to the two Weyl points, the corresponding Lagrangian reads  $\mathcal{L}=\bar{\psi}(i\slashed\partial-\slashed A-\gamma^{5}\slashed b)\psi$, where $\mathbf{b}$ is promoted  to  be a space-time vector $b_\mu$ with $b_0$ corresponding to deviation in energy, and notations about Dirac matrices are consistent with those in \cite{Peskin-Book}. There are two types of vertexes, \chiralvex$=-\int d^4x\bar{\psi}\gamma^5\slashed b\psi$ and \photonvex$=-\int d^4x\bar{\psi}\slashed A\psi$. The A-CS term is given by the three-vertex diagrams with two \photonvex's and one \chiralvex, shown in Fig.(\ref{Feynman}). As the $U(1)$ gauge symmetry is fundamental, we adopt the dimensional regularization scheme for the internal momentum $l$ in the loop. Accordingly $l$ is decomposed as $\slashed l=\slashed l_{\parallel}+\slashed l_{\perp}$.
In the above notations, $\slashed l_{\parallel}$ is still in the four-dimensional
spacetime, but $\slashed l_{\perp}$ is in the extended infinitesimal dimensions, anti-commuting with $\gamma^{\mu}$ and commuting with $\gamma^{5}$~\cite{Hooft,Peskin-Book}. The three diagrams contribute equally, and after a dramatic cancellation only a finite term is left similar to the derivation of the axial anomaly~\cite{Anomaly1,Anomaly2,Peskin-Book}, leading to the anisotropic CS term,
\begin{equation}
 S_{ACS}=-\frac{1}{4\pi^{2}}\int d^{4}x\:\epsilon^{\mu\nu\rho\sigma}b_{\mu}A_{\nu}\partial_{\rho}A_{\sigma}(x).\label{ACS}
\end{equation}

\textit{Dimension reduction} The last arrow process to be established is the dimension reduction from $(2n+1)$D CSM to $(2n)$D TI in class A. First it is observed that if $\mathbf{b}$ is a constant vector in the model (\ref{Model}), along the direction of $\mathbf{b}$, we may regard the 3D system as a collection of 2D systems perpendicular to $\mathbf{b}$ in $\mathbf{k}$ space, and thus the 2D systems gapped over $\mathbf{k}\in(-\mathbf{b},\mathbf{b})$ are 2D TIs with unit Chern number and the others outside this range are trivial, except at the two gapless points. Setting $\mathbf{b}=(0,0,b)$ constant and all fields independent of $z$ in Eqs.(\ref{ATheta}) and (\ref{ACS}), the dimension reduction gives
$
S_{A\Theta}=-\frac{bL_z}{8\pi}\int d^2 x\,\,\epsilon^{jk}\mathrm{tr}(Q\partial_jQ\partial_kQ)
$
and
$
S_{ACS}=-\frac{bL_z}{4\pi^2}\int d^3 x\,\,\epsilon^{\nu\rho\sigma}A_\nu \partial_\rho A_\sigma.
$
Since $k_z\in (-b,b)$ correspond to a collection of unit topological insulators (Chern insulators), divided by the dimension constant $2b\times \frac{L_z}{2\pi}$, the above equations give the well-known $\theta$-term and CS term for 2D TIs,
\begin{eqnarray}
 S_{\Theta}^{2D}&=&-\frac{1}{8}\int d^2 x\,\, \epsilon^{jk} \mathrm{tr}(Q\partial_j Q\partial_k Q)\label{2DTheta}\\
  S_{CS}^{2D}&=&-\frac{1}{4\pi}\int d^{3}x\:\epsilon^{\mu\nu\rho}A_{\mu}\partial_{\nu}A_{\rho}(x)\label{2DCS}.
\end{eqnarray}
 Since the above treatment is applicable for any integer $n>0$, the dimension reduction(DR) from $(2n+1)$D CSM to $(2n)$D TI in Fig.(\ref{diagram}) has been completed. The dimension reduction actually illustrates the correspondence between the anisotropic $\theta$-term of Eq.(\ref{ATheta}), and the anisotropic CS term of Eq.(\ref{ACS}), recalling that the CS term for 2D TI indicates the transverse conductivity and the $\theta$-term implies the stability of this transverse transportation under disorders. 

\textit{Remark}  We now make comments about the essence of emergent topological terms. In quantum field theory, this emergence 
is usually related to some quantum anomaly, where two regularization schemes with different symmetries are contradictory~\cite{Anomaly1,Anomaly2,Pruisken}. From a viewpoint of band theory,
these contradictions are usually originated from nontrivial topological configurations of the Berry fiber bundle. The coupling constant of a topological term can formally be expressed as a topological invariant of the Berry fiber bundle, 
which
affects directly the transport properties~\cite{BerryRMP}, and thus its nontrivial topological configurations may account for anomalous transport in quantum anomalies. 

\begin{acknowledgments}
This work was supported by the GRF (Grant Nos. HKU7045/13P and HKU173051/14P) and the CRF (HKU8/11G) of Hong Kong.
\end{acknowledgments}

\section{Supplemental Material}


In this supplemental material, we present the derivation details of the anisotropic Chern-Simons term through opposite coupling. As mentioned in the main text, it comes from the summation of the three diagrams with one \chiralvex\,and two \photonvex's. Let $\mathbf{III}$ denote the summation. Applying Feynman rules we have
\begin{eqnarray*}
\mathbf{III}&=&-\int\frac{dpdkdl}{(2\pi)^{4\times3}}\mathrm{tr}[(-i\gamma^{5}\slashed b(-p-k))\frac{i(\slashed l-\slashed k)}{(l-k)^{2}}(-i\slashed A(k))\\
&&~~~~~\frac{i\slashed l}{l^{2}}(-i\slashed A(p))\frac{i(\slashed l+\slashed p)}{(l+p)^{2}}],
\end{eqnarray*}
noting that the three diagrams are equal. Then the above express can
be rewritten as
\[
\mathbf{III}=-\int\frac{dpdk}{(2\pi)^{8}}(-ib_{\mu}(-p-k))A_{\lambda}(k)A_{\nu}(p)\mathcal{M}^{\mu\nu\lambda}(p,k),
\]
where
\[
\mathcal{M}^{\mu\nu\lambda}(p,k)=(-i)^{2}\int\frac{d^{4}l}{(2\pi)^{4}}\mathrm{tr}\left[\gamma^{5}\gamma^{\mu}\frac{i(\slashed l-\slashed k)}{(l-k)^{2}}\gamma^{\lambda}\frac{i\slashed l}{l^{2}}\gamma^{\nu}\frac{i(\slashed l+\slashed p)}{(l+p)^{2}}\right].
\]
Introducing Feynman's parameters $x$ and $y$ and replacing $l$
by $l+xk-yp$, the above expression can be simplified as
\begin{eqnarray*}
\mathcal{M}^{\mu\nu\lambda}(p,k) & = & -2i\int_{0}^{1}dxdy\int\frac{d^{d}l}{(2\pi)^{d}}\frac{1}{(l^{2}-\Delta)^{3}}\\
 &  & \mathrm{tr}[\gamma^{5}\gamma^{\mu}(\slashed l+(x-1)\slashed k-y\slashed p)\gamma^{\lambda}(\slashed l+x\slashed k-y\slashed p)\\
 & & \gamma^{\nu}(\slashed l+x\slashed k+(1-y)\slashed p)].
\end{eqnarray*}
As the $U(1)$ gauge symmetry is fundamental, we adopt the dimensional
regularization scheme for $l$. Accordingly, $l$ is decomposed as
\[
\slashed l=\slashed l_{\parallel}+\slashed l_{\perp}.
\]
In the above notation, $\slashed l_{\parallel}$ is still in the four-dimensional
spacetime, but $\slashed l_{\perp}$ is in the extended spacetime,
anti-commuting with $\gamma^{\mu}$ and commuting with $\gamma^{5}$.
It is noted that $p$ and $k$ are still in the four-dimensional spacetime
without extension. The only non-vanishing terms within the trace
are
\[
\begin{array}{cc}
 & \mathrm{tr}\left\{ \gamma^{5}\gamma^{\mu}[(x-1)\slashed k-y\slashed p]\gamma^{\lambda}\slashed l_{\perp}\gamma^{\nu}\slashed l_{\perp}\right\} \\
+ & \mathrm{tr}\left\{ \gamma^{5}\gamma^{\mu}\slashed l_{\perp}\gamma^{\lambda}\slashed l_{\perp}\gamma^{\nu}[x\slashed k+(1-y)\slashed p]\right\} \\
+ & \mathrm{tr}\left\{ \gamma^{5}\gamma^{\mu}\slashed l_{\perp}\gamma^{\lambda}(x\slashed k-y\slashed p)\gamma^{\nu}\slashed l_{\perp}\right\} .
\end{array}
\]
Their summation can be simplified as
\[
4i\epsilon^{\mu\lambda\nu\alpha}[(x-1)k_{\alpha}+(1-y)p_{\alpha}]l_{\perp}^{2},
\]
using the identity
\[
\mathrm{tr}(\gamma^{5}\gamma^{\mu}\gamma^{\nu}\gamma^{\rho}\gamma^{\sigma})=-4i\epsilon^{\mu\nu\rho\sigma}.
\]
Thus
\begin{eqnarray*}
\mathcal{M}^{\mu\nu\lambda}(p,k) & = & 8\epsilon^{\mu\lambda\nu\alpha}\int_{0}^{1}dxdy[(x-1)k_{\alpha}+(1-y)p_{\alpha}]\\
&&\int\frac{d^{d}l}{(2\pi)^{d}}\frac{l_{\perp}^{2}}{(l^{2}-\Delta)^{3}}\\
 & = & -4\epsilon^{\mu\lambda\nu\alpha}(k_{\alpha}-p_{\alpha})\int\frac{d^{d}l}{(2\pi)^{d}}\frac{l_{\perp}^{2}}{(l^{2}-\Delta)^{3}}.
\end{eqnarray*}
Using $l_{\perp}^{2}=\frac{d-4}{d}l^{2}$, the integration above is
evaluated as
\begin{eqnarray*}
\int\frac{d^{d}l}{(2\pi)^{d}}\frac{l_{\perp}^{2}}{(l^{2}-\Delta)^{3}}&=&\frac{i}{(4\pi)^{d/2}}\frac{d-4}{2}\frac{\Gamma(2-\frac{d}{2})}{\Gamma(3)\Delta^{2-d/2}}\\
&\underset{_{d\rightarrow4}}{\longrightarrow}&\frac{-i}{2(4\pi)^{2}}.
\end{eqnarray*}
So we have
\[
\mathcal{M}^{\mu\nu\lambda}(p,k)=\frac{2i}{(4\pi)^{2}}e^{\mu\lambda\nu\alpha}(k_{\alpha}-p_{\alpha}).
\]
Now we are ready to evaluate $\mathbf{III}$ as
\begin{eqnarray*}
\mathbf{III}& = & -\frac{2}{(4\pi)^{2}}\int d^{4}x\;\epsilon^{\mu\lambda\nu\alpha}(b_{\mu}(x)i\partial_{\alpha}A_{\lambda}(k)A_{\nu}(p)\\
&&-b_{\mu}(x)A_{\lambda}(k)i\partial_{\alpha}A_{\nu}(p))\\
 & = & -i\frac{1}{4\pi^{2}}\int d^{4}x\;\epsilon^{\mu\nu\rho\sigma}b_{\mu}A_{\nu}\partial_{\rho}A_{\sigma}(x).
\end{eqnarray*}
Thus the effective theory also has the anisotropic Chern-Simons term
\[
S_{ACS}=-\frac{1}{4\pi^{2}}\int d^{4}x\:\epsilon^{\mu\nu\rho\sigma}b_{\mu}A_{\nu}\partial_{\rho}A_{\sigma}(x).
\]


\begin{thebibliography}{References}
\bibitem{note0} Although the displacements of two Weyl points are assumed to be placed oppositely in the momentum space due to IS for simplicity, our main results are valid for a general case without IS.
\bibitem{note1} Here the term `chiral point' refers to a gapless point in any odd dimensional momentum space such that the topological charge
is nontrivial.
\end{thebibliography}
\end{document}